\documentstyle[aps,prd,preprint,psfig]{revtex}
\newcommand{\be}{\begin{equation}}
\newcommand{\ee}{\end{equation}}
\newcommand{\ben}{\begin{eqnarray}}
\newcommand{\een}{\end{eqnarray}}   
\begin{document}

\title{CHARGED VACUUM CONDENSATE NEAR A SUPERCONDUCTING COSMIC
STRING}

\author{J.R.S. Nascimento\footnote{Electronic Address: jroberto@cosmos2.phy.
tufts.edu}
\thanks{On leave from Departamento de F\'\i sica,
Universidade Federal da Para\'\i ba, Caixa Postal 5008, 58051-970 Jo\~ao
Pessoa, Para\'\i ba, Brazil}, Inyong Cho\footnote{Electronic Address:
cho@cosmos2.phy.tufts.edu} and Alexander Vilenkin\footnote{Electronic Address:
vilenkin@cosmos2.phy.tufts.edu}}

\address{Institute of Cosmology, Department of Physics and Astronomy,\\
Tufts University, Medford, Massachusetts 02155}

\date{\today}

\maketitle

\begin{abstract}

A charged superconductiong cosmic string produces an extremely large
electric field in its vicinity.  This leads to vacuum instability and
to the formation of a charged vacuum condensate which screens the
electric charge of the string.  We analyze the structure of this
condensate using the Thomas-Fermi method.
\end{abstract}
\vspace{0.3cm}
PACS number(s): 98.80.Cq

\newpage

\section{INTRODUCTION}
Cosmic strings are linear defects that could be formed at a
phase transition in the early universe. (For a review see\cite{VI}.)
Witten\cite{EW} has shown that 
strings predicted in some grand unified models behave as 
superconducting wires.  Such strings moving through magnetized cosmic
plasmas can develop large currents and can give rise to a variety of
astrophysical effects.  In particular, they have been suggested as
possible sources of ultrahigh energy cosmic rays \cite{HSW}.  

Currents developed by oscillating strings in a magnetic field are not
homogeneous along the strings because different portions of the string
cross the magnetic field lines in different directions.  This results
in charge accumulation, and portions of the string can develop a charge
per unit length $\lambda$ comparable to the current, $\lambda\sim J$.
(Here and below we use units in which $\hbar=c=1$.)
The electric field near the string is given by
\be
\label{evac}
E = \frac{2\lambda}{r}.
\ee
It can become extremely strong in the immediate vicinity of the string, and
then quantum effects, such as vaccum polarization and pair production, must 
be taken into account.  One can expect that the created particles of
charge opposite to that of the string will accumulate in bound states
and form a condensate screening the electric field near the string to 
below the critical
value.  It was noted earlier that such screening would lead to a
drastic modification of string electrodynamics \cite{Aryal} and would
have a significant effect on the propagation of high-energy particles
emitted from the charged portions of the string \cite{BR}.  The purpose
of the present paper is to give a quantitative description of the
screening condensate near a charged superconducting string.

The superconducting current in the strings
is carried by charged particles which acquire a mass $M$ at the
string-forming phase transition but remain massless inside the
strings.  These massless charge carriers move along the strings 
at the speed of light.  The string current is bounded by the critical
value, 
\be
J_c\sim eM, 
\label{Jc}
\ee
at which the characteristic energy of the charge
carriers becomes comparable to $M$, so that they have enough energy to
jump out of the string.  The mass $M$ is model-dependent but is
limited by the string symmetry breaking scale $\eta$, $M\lesssim\eta$.

In a
cosmological setting, the string charges and currents vary on
astronomical time and length scales, and for our purposes we can
regard them as constant.
For a string segment with
$J<\lambda$, we can always find a Lorentz frame where $J=0$.   The
electric field close to the string will then be well approximated by
that of an infinite straight string. 
We shall consider, therefore, an infinite straight string with a
constant charge per unit length
$\lambda$ and vanishing current, $J=0$.  For sufficiently large
$\lambda$, charged particles (for definiteness electrons) have bound
states localized near the string with negative energies smaller than
$-m$, where $m$ is the electron mass.  The vacuum then becomes
unstable with respect to production of electron-positron pairs.  For a
positively charged string, positrons are repelled away, while
electrons form a vacuum condensate surrounding the string.  We shall
determine the electric field and the charge distribution in this
condensate using the Thomas-Fermi method \cite{TF} in which the
condensate is 
approximately treated as an ideal gas obeying the Fermi-Dirac
statistics. 

In the next section we shall review the derivation of the relativistic
Thomas-Fermi equation and specify the boundary conditions appropriate 
for the case of cylindrical symmetry.  Approximate analytic solutions
of this equation are given in Sec. III, and its numerical solutions
are presented in Sec. IV.  
The conclusions of the paper are summarized 
and discussed in Sec. V.

\section{Thomas-Fermi equation}

The density of electrons in a degenerate Fermi gas is related to the
Fermi momentum $p_{F}$ by
\be
\label{n}
n_{e}(r) = \frac{p_{F}^3}{3\pi^2}.
\ee
The relativistic relation between the Fermi energy $\epsilon_{F}$ and Fermi 
momentum is
\be
\label{p}
p_{F} = \left[(\epsilon_{F} - V(r))^2 - m^2\right]^{1/2}
\ee   
where $m$ is the electron mass, $-e$ is its charge, $V(r)=
-e\varphi(r)$ and $\varphi(r)$ is the 
self-consistent electrostatic potential for an electron,
taking into account
both the field of the string and the average field produced by other
electrons of the condensate.
The condensate is formed of electrons occupying quantum states in the
negative energy continuum, $\epsilon < -m$. We therefore set the 
fermi energy to be  $\epsilon_{F}= -m$.
The electron density (\ref{n}) is then given by
\be
\label{n1}
n_{e}(r)= \frac{1}{3\pi^2} \left[V^2(r)+2mV(r)\right]^{3/2}
\ee

Introducing the total charge density $\rho_{T}$ which is composed of the 
electron charge and external string charge,
\be
\rho_{T}=\rho_{s}- en_{e}
\ee
and using the Poisson equation
\be
\triangle V(r) = 4\pi e \rho_{T}(r),
\label{poisson}
\ee
we find a self-consistent non-linear differential
equation
\ben
\label{pe}
\triangle V(r) =-4\pi e\left[\frac{e}{3\pi^2}\left(V^2(r) + 
2mV(r)\right)^{3/2} 
-\rho_{s}(r)\right].
\een
This equation has been used in \cite{MI76,BJ} to study
the electron condensate around supercharged nuclei.  In the case of a
string, the problem has cylindrical symmetry and
\be
\triangle V(r) = V''(r) +{1\over{r}}V'(r). 
\ee

We shall approximate the string charge distribuition as a uniform 
distribuition in  a cylinder of radius $\delta$, 
\be
\rho_{s}(r)=\rho_{0}\theta(\delta-r).
\ee
The linear charge 
density of the string is given by $\lambda=\pi \delta^2 \rho_0$.  The
charge carriers are typically concentrated in a tube of radius $r\sim
M^{-1}$; hence, we should have $\delta\sim M^{-1}$.

It is easily seen from Eq. (\ref{n1}) that the density of electrons,
$n_{e}(r)$, is different from zero only in the 
region of space where $V(r)<-2m$. Therefore, the condensate has a 
finite radius $r=R_{c}$. For $r>R_c$, the solution of (\ref{pe}) is
just the usual logarithmic 
potential of a linear charge,
\be
V(r) = 2e\lambda_0\ln\frac{r}{R_{\ast}}
\ee
Here,  $\lambda_0$ is the total charge per unit length 
of string, including both the charge carriers in the core and the condensate,
and $R_{\ast}$ is the cutoff radius indicating the distance at which 
the approximation of an 
infinite straight string breaks down. $R_{\ast}$ is given by the smallest 
of the following three length scales: (i) the typical distance between 
the strings in a cosmic string network, (ii) the characteristic curvature 
radius of string, (iii) the typical wavelength of the current-charge 
oscillations along the string. 
 
The boundary condition for Eq. (\ref{pe})
at $r=0$ is 
\be
\label{c1}
V'(0)=0,
\ee
while at $r=R_{c}$ we have
\be
\label{c2}
V(R_{c})=-2m,~~~~~~~~~~~ V'(R_{c})= 
\frac{2}{R_{c}\ln(R_{\ast}/R_{c})}.
\ee
Note that we have three rather than two boundary conditions, as a second-order 
differential equation would normally require. The third condition is needed 
 to determine the condensate radius $R_c$

We expect $R_c$ to be microscopic, while $R_{\ast}$ will typically be astrophysically large. Hence, the logarithm in Eq.(\ref{c2}) is 
$\ln(R_{\ast}/{R_c}) \sim 10^2$. In numerical calculations below we choose 
$R_{\ast}$ so that 
$\ln({R_{\ast}}/{R_c}) \approx 30$; our results are not sensitive to this choice.

The Thomas-Fermi approximation is adequate when the characteristic
scale of variation of the condensate density $n_e(r)$ is large
compared to the electron wavelength $1/p(r)$.  The corresponding
condition is
\be
\label{c}
\left| \frac{d}{dr}\left[\frac{1}{p(r)}\right]\right| \ll 1.
\ee
We shall see that this condition is satisfied in most of the
condensate region $0<r<R_c$, provided that the charge density
$\lambda$ is sufficiently large.

\section{Analytic approximations}

The Thomas-Fermi equation (\ref{pe}) can be solved analytically in the
limit when the magnitude of the potential $V(r)$ is large, $\left|V(r)\right|\gg 2m$.  
We can then
neglect $2mV(r)$ compared to $V^2(r)$, and outside the string core
Eq.(\ref{pe}) reduces to
\be
V''(r)+{1\over{r}}V'(r)=-{4e^2\over{3\pi}}|V(r)|^3.
\ee
This has a solution
\be
V(r)=-C/r
\label{va}
\ee
with
\be
C=(3\pi/4e^2)^{1/2}\approx 18.
\label{ca}
\ee
The corresponding electric field is
\be
E(r)=C/er^2.
\label{ea}
\ee

We note that the solutions (\ref{va}) and (\ref{ea}) do not depend on the
string charge density $\lambda$.  As $r$ decreases, the electric field
(\ref{ea}) grows faster than that of the vacuum solution (\ref{evac}).
It cannot, therefore, be extended all the way to the string but has to
be matched with Eq.~(\ref{evac}) at some radius $R_s$ below which the
vacuum solution takes over.  The matching radius at which the two
electric fields become comparable is 
\be
R_s\sim C/e\lambda\sim 200\lambda^{-1}.
\label{rs}
\ee
We shall call it the screening radius.  For $r\ll R_s$, the screening
is unimportant and the electric field is given by Eq.~(\ref{evac}).  
The screening radius is always large compared to the string thickness
$\delta\sim M^{-1}$, provided that $\lambda$ is smaller than the
critical value (\ref{Jc}), 
\be
\lambda\lesssim eM.
\label{lc}
\ee

The potential corresponding to the vacuum solution (\ref{evac}) at
$\delta<r\ll R_s$ is
\be
V(r)= -2e\lambda[\ln (R_s/r)+B],
\ee
where $B\sim 1$ is a numerical constant.  The potential at the string
core is thus
\be
V(0)\approx -2e\lambda\ln(R_s/\delta).
\label{v0}
\ee

The condition $\left|V(r)\right|\gg m$  implies $r\ll
C/m$, and thus  the solution (\ref{ea}) is valid in the range
$C/e\lambda \ll r\ll C/m$.  This range exists only if $\lambda$ is
sufficiently large, $\lambda\gg m/e$.  Combined with the condition
(\ref{lc}) this implies $M\gg m/e^2$.  In models of astrophysical
interest, the charge carrier mass $M$ is very large (so that the
strings can develop large currents and charges), and this condition is
satisfied with a large margin.

At $r\sim C/m$, the potential $V(r)$ becomes comparable to $-m$
signalling that we are close to the condensate boundary [see Eq.~(\ref{c2})].
Hence, we can estimate the condensate radius as 
\be
R_c \sim C/m.
\label{rc}
\ee

The condition of validity of the Thomas-Fermi approximation (\ref{c}),
when applied to the solution (\ref{va}), gives $C\gg 1$.  This is
satisfied with a reasonable accuracy [see Eq.~(\ref{ca})].

\section{NUMERICAL CALCULATION}

We obtained numerical solutions to the Thomas-Fermi equation for 
Eq.~(\ref{pe}) for $V(r)$ with the boundary conditions 
(\ref{c1}) and (\ref{c2}) using the relaxation method.  
The resulting electric field is plotted in Fig. 1, together with the
analytic approximations (\ref{evac}) and (\ref{ea}).  The agreement
between the analytic and numerical solutions is excellent in the
appropriate ranges of the radius $r$.  

We have verified that the shape of $V(r)$ outside the 
string core is not sensitive to the value of the core radius
$\delta$.  In particular, the 
condensate radius $R_c$ approaches a constant value independent of $\delta$
 (see  Fig.~\ref{fig=spRaRc}). This is very fortunate, since a realistic value
 of the core radius would be too small to resolve in our calculations. 
Figure~\ref{fig=spRaRc} suggests that 
it is sufficient to choose $\delta \ll m^{-1}$. We used $\delta =
10^{-3}m^{-1}$ in most
of the calculations described below.

The condensate radius $R_c$ is plotted in Fig.~\ref{fig=spRaZ},
as a function of the linear charge density of the string, $\lambda$. 
We see that at large $\lambda$, $R_c$ approches a constant value,
\be
R_c = 82 m^{-1},
\ee
in agreement with Eq.~(\ref{rc}).
The screened linear charge density of the string $\lambda_0$, which 
determines the electric field outside $R_c$, is also found to be
independent of $\lambda$:
\be
\lambda_0 \approx 5.34 em.
\ee
The electric field at the condesate boundary is
\be
E_0=2\lambda_0/R_c \approx 10^{-1}em^2.
\ee
Note that this is considerably smaller than the critical field,
$E_c=m^2/e$, which signals the onset of intensive pair production
\cite{Schwinger}.  In our case, $E_0\approx 10^{-3}E_c$.  We
shall return to this point later in Sec. V.

The effective linear charge density $\lambda_{eff}(r)$ inside the condensate 
can be found as
\be
\lambda_{eff}(r) = 2\pi\int_0^r\rho(r')r'dr'=\frac{r}{2e}\frac{dV}{dr}
\ee
where we have used Eq.~(\ref{poisson}). 
As $r$ grows, $\lambda_{eff}(r)$ decreases and we can 
define the effective screening radius $R_s$ as the radius at which
half of the string charge is screened,
\be
\lambda_{eff}(R_s) = \lambda/2.
\ee
At the boundary of the condensate we must have 
$\lambda_{eff}(R_c) = \lambda_0$. The screening radius $R_s$ is plotted
in Fig.~4 for several values of $\lambda$. We see that, although the condensate
radius $R_c$ is independent of $\lambda$, the screening radius gets
smaller rather rapidly as $\lambda$ is increased.
A numerical fit to the data in the Fig.~4 gives
\be
R_s = 80 \lambda^{-1},
\ee
in agreement with the order-of-magnitude estimate (\ref{rs}).

\section{Summary and DISCUSSION}

We have found that a superconducting cosmic string having a
sufficiently large charge per unit length, $\lambda\gg m/e$, is
surrounded by an electron condensate of radius $R_c\sim 100m^{-1}$,
where $m$ is the electron mass.  In the immediate vicinity of the
string, the effect of the condensate is unimportant and the electric
field is given by the vacuum solution, $E\approx 2\lambda/r$.  Screening due
to the condensate becomes significant at $r\sim R_s \sim
100\lambda^{-1}$, and for $R_s\ll r\ll R_c$ the electric field has the
form $E\approx (3\pi)^{1/2}/2e^2r^2$.  Outside the condensate, at
$r>R_c$, the field is given by $E\approx em/r\approx
10^{-2}em^2(R_c/r)$. 

As we already mentioned, the electric field at the condensate boundary
is well below the critical field, $E_0\sim 10^{-3}E_c$, where
$E_c=m^2/e$.  
The rate of pair production per unit volume in a homogeneous electric field is
 \cite{Schwinger} 
\be
dN/dVdt \approx (eE/\pi)^2\exp(-\pi E_c/E),
\ee
which indicates that the outer parts of the condensate where $E\ll
E_c$ will be filled up very slowly.  The characteristic time of pair
production, $\tau\sim (eE)^{-1/2}\exp(\pi E_c/E)$, is greater than
the age of the universe for $E \lesssim 4\times 10^2 E_c$.  For
astrophysical strings, we expect the condensate radius to be given by
the distance from the string at which such values of the electric
field are reached.  From
Eq.~(\ref{va}) we find 
\be
R_c\sim {\sqrt{C}\over{20}}m^{-1}\approx 0.2 m^{-1}.
\ee


As the charge density of the string $\lambda$ is increased, the
potential near the string core becomes more and more negative.  As a
result, particles more massive than electrons develop condensates.
From Eq.~(\ref{v0}), particles of mass $\mu$ develop states with
$\epsilon <-\mu$ at $\lambda\sim \mu/e\ln(R_s/\delta)\sim\mu$.  The
condensates of different particle species will have the form of
coaxial cylinders, with condensates of more
massive particles being closer to the string.

Finally, we would like to mention some open questions.  In this paper
we studied vacuum condensation of fermions.  Charged Bose particles,
such as Higgs and gauge bosons will also form vacuum condensates, and
the properties of these bosonic condensates may differ from the
fermionic case.  Another important problem is the nature of
modifications introduced by vacuum screening in string electrodynamics
and in the propagation of charged particles emitted by the strings.
We hope to return to some of these issues in future publications.

\acknowledgments
{J.R.S.N. is grateful to the Institute
of Cosmology, Tufts University, for hospitality.
The work of J.R.S.N. was supported in part by funds provided by
Conselho Nacional Desenvolvimento Cient\'\i fico e Tecnol\'ogico,
CNPq, Brazil.  The work of I.C. and A.V. was supported in part by the National
Science Foundation.}

\begin{figure}
\psfig{file=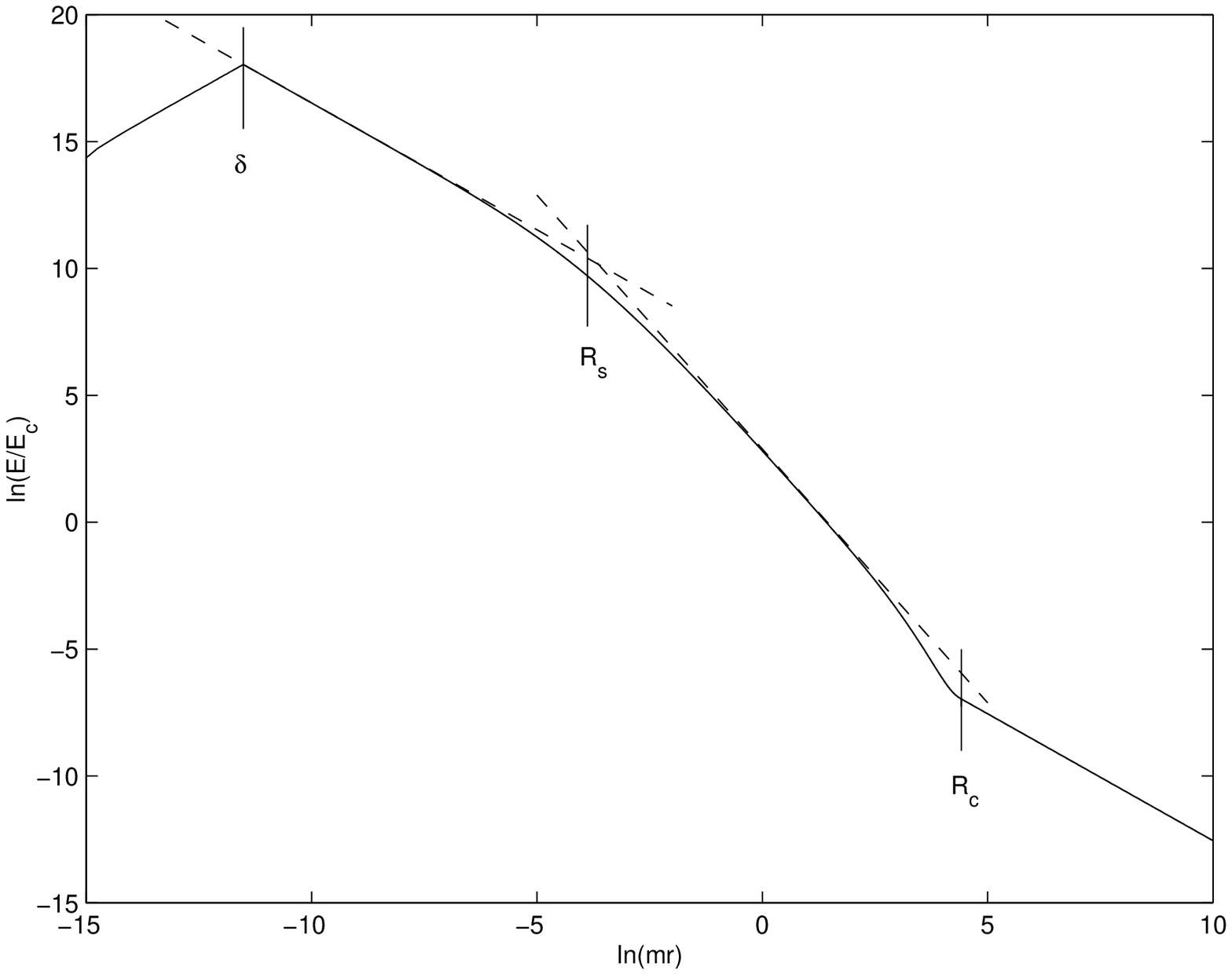}
\caption{The electric field $E$ in units of the critical field
$E_c=m^2/e$ is shown as a function of the distance from the string $r$
for $\lambda=4\times 10^3 m$ and $\delta=10^{-5}m^{-1}$ (solid line).  
Dotted lines indicate the analytic approximations
(\ref{evac}) and (\ref{ea}) in the appropriate regimes.  The core
radius $\delta$, the screening radius $R_s$ and the condensate radius
$R_c$ are also indicated.}

\label{fig=spVRc}
\end{figure}

\begin{figure}
\psfig{file=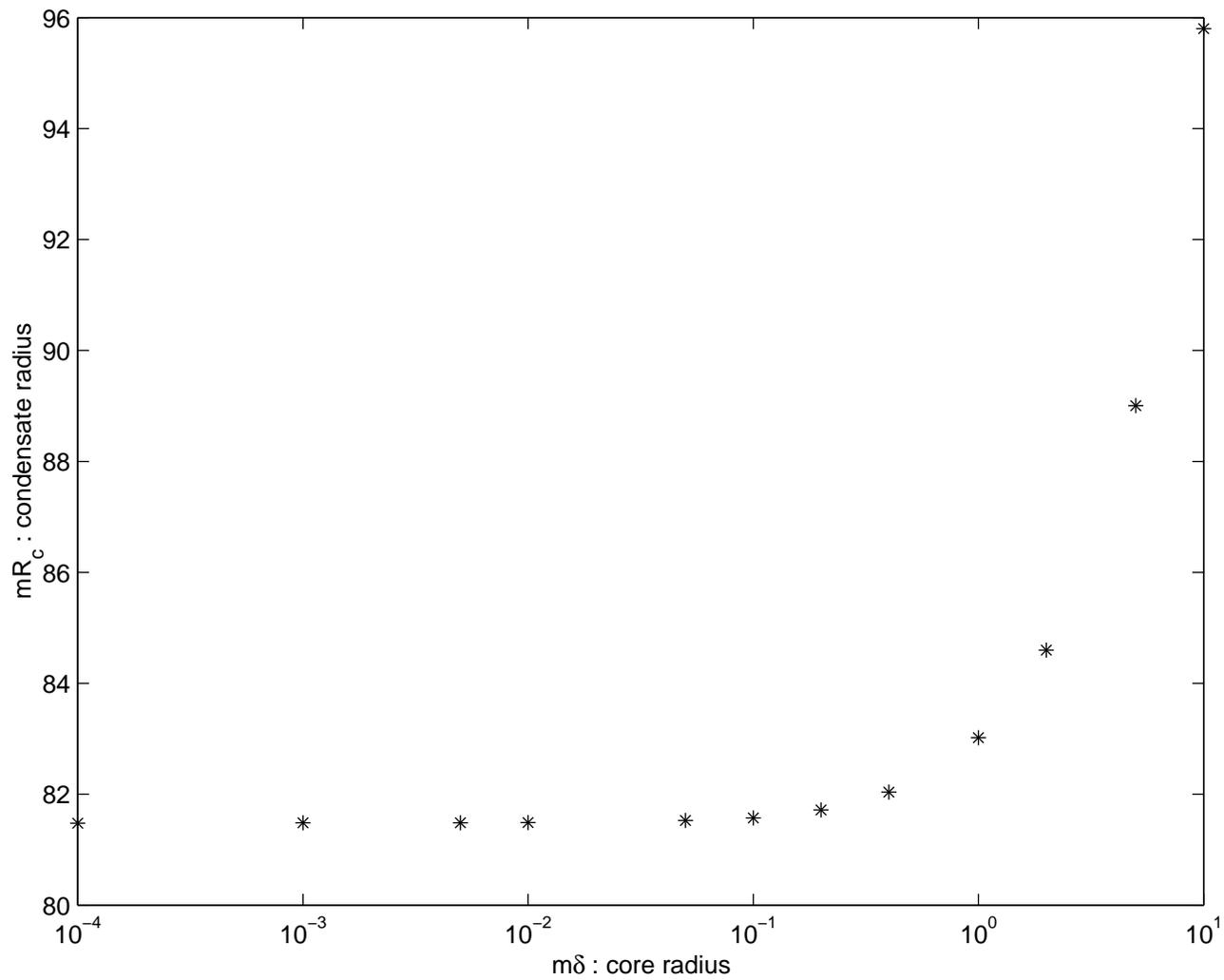}
\caption{The condensate radius $R_c$ vs the core radius $\delta$ 
for $\lambda=270 m^{-1}$.} 
\label{fig=spRaRc}
\end{figure}

\begin{figure}
\psfig{file=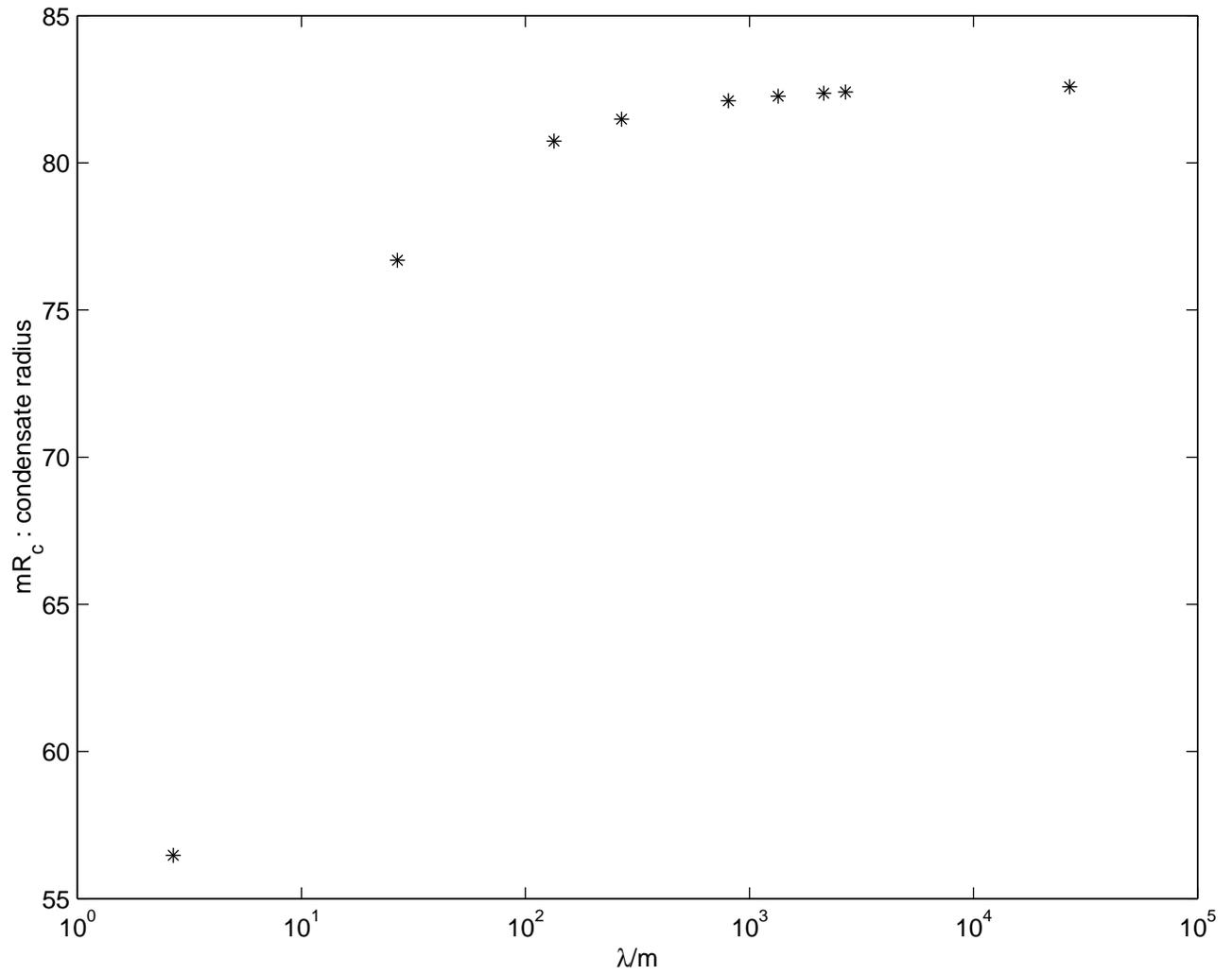}
\caption{The condensate radius $R_c$ vs the linear charge
density of the string, $\lambda$ for $\delta =0.001m^{-1}$.} 
\label{fig=spRaZ}
\end{figure}

\begin{figure}
\psfig{file=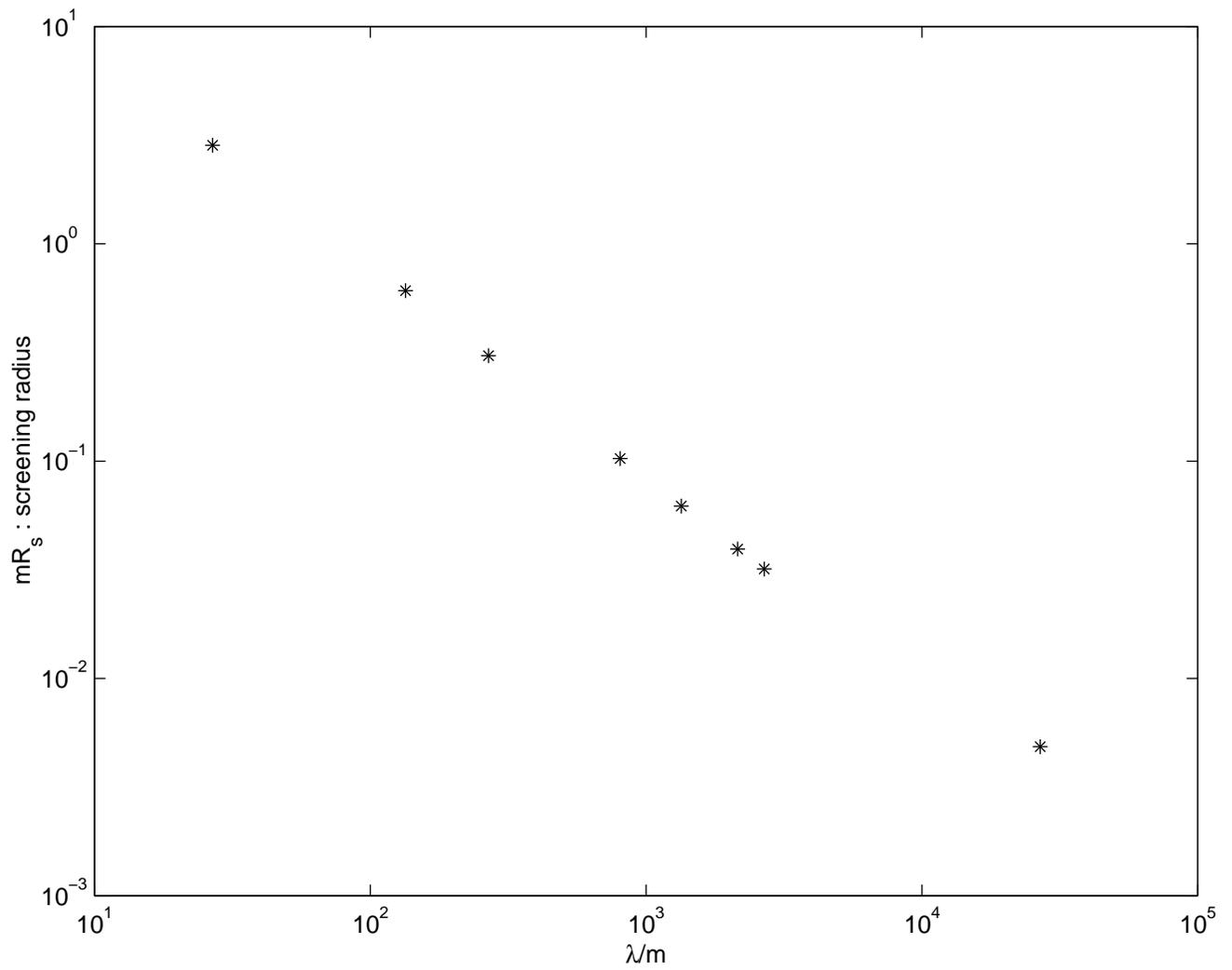}
\caption{The screening radius $R_s$ vs the linear charge
density of the string $\lambda$ for $\delta =0.001m^{-1}$.}
\label{fig=spRsZ}
\end{figure}
\end{document}